\documentclass[12pt,preprint]{aastex}
\usepackage{graphicx}
\usepackage{lscape}

\usepackage{spr-astr-addons}
\usepackage{natbib}
\usepackage{graphicx}

\begin{document}

\title{Pulse phase-resolved analysis of SMC X-3 during its 2016--2017 super-Eddington outburst}

\shorttitle{Pulse phase-resolved analysis of SMC X-3}
\shortauthors{Zhao et al.}

\author{Hai-Hui Zhao\altaffilmark{1}, Shan-Shan Weng\altaffilmark{1}, Ming-Yu Ge\altaffilmark{2}, Wei-Hao Bian\altaffilmark{1}, Qi-Rong Yuan\altaffilmark{1}}
\email{zhaohh@njnu.edu.cn, wengss@njnu.edu.cn}

\altaffiltext{1}{Department of Physics and Institute of Theoretical Physics,
Nanjing Normal University, Nanjing 210023, China; wengss@njnu.edu.cn}

\altaffiltext{2}{Key Laboratory of Particle Astrophysics, Institute of High
Energy Physics, Chinese Academy of Sciences, Beijing 100049, China}

\begin{abstract}
The Be X-ray pulsar SMC X-3 underwent an extra long and ultraluminous
giant outburst from 2016 August to 2017 March. The peak X-ray luminosity is up
to $\sim 10^{39}$ erg/s, suggesting a mildly super-Eddington accretion onto the
strongly magnetized neutron star. It therefore bridges the gap between the
Galactic Be/X-ray binaries ($L_{\rm X}^{\rm peak} \leq 10^{38}$ erg/s) and the
ultraluminous X-ray pulsars ($L_{\rm X}^{\rm peak} \geq 10^{40}$ erg/s) found
in nearby galaxies. A number of observations were carried out to observe the
outburst. In this paper, we perform a comprehensive phase-resolved
analysis on the high quality data obtained with the {\it Nustar} and {\it
XMM-Newton}, which were observed at a high and intermediate luminosity levels.
In order to get a better understanding on the evolution of the whole extreme
burst, we take the {\it Swift} results at the low luminosity state into account
as well. At the early stage of outburst, the source shows a double-peak pulse
profile, the second main peak approaches the first one and merges into the
single peak at the low luminosity. The second main peak vanishes beyond 20 keV,
and its radiation becomes much softer than that of the first main peak. The
line widths of fluorescent iron line vary dramatically with phases, indicating
a complicated geometry of accretion flows. In contrast to the case at low
luminosity, the pulse fraction increases with the photon energy. The
significant small pulse fraction detected below 1 keV can be interpreted as the
existence of an additional thermal component located at far away from the
central neutron star.

\end{abstract}

\keywords{accretion, accretion disks --- stars: neutron --- pulsars: general
--- X-rays: binaries  --- X-rays: individual (SMC X-3)}

\section{Introduction}

A Be/X-ray binary (BeXRB) consists of a Be star and a compact object.
Only a few sources are identified as an accreting black-hole
\citep[e.g., MWC 656, ][]{casares14} or perhaps a white-dwarf \citep[e.g.,
$\gamma$ Cassiopeia, ][]{haberl95, postnov17}, while most of confirmed
compact objects in BeXRBs are neutron stars \citep[NSs; see ][for
reviews]{bildsten97, reig11}, and half of these systems show X-ray pulsations
\citep{haberl16}. BeXRBs spend most time at the quiescence state, interrupted
by quasi-periodic and less energetic outbursts or rare giant outbursts, i.e.
type I and type II outbursts, respectively. A pulse phase-resolved analysis on
the type II outburst is essential to test the accretion theory on strongly
magnetized NSs at different accretion states.

The BeXRB SMC X-3 was discovered with the {\it SAS-3} X-ray
observatory in Small Magellanic Cloud \citep{clark78}, and its pulsation ($P =
7.78$ s) was detected by other X-ray missions, e.g., {\it Chandra}
\citep{edge04}, {\it XMM-Newton} \citep{haberl08}, and {\it RXTE}
\citep{galache08}. From 2016 August to 2017 March, SMC X-3 underwent a
super-Eddington outburst with a peak luminosity of $\sim 10^{39}$ erg/s, making
it as the most luminous giant outburst reported in BeXRBs
\citep{weng16, weng17, townsend17, tsygankov17}. Investigating the follow-up
{\it Swift} monitoring observations, we found that the pulse profile exhibited
a double-peak profile at the high luminosity state and then merging into the
single peak at the low luminosity \citep[][hereafter Paper I]{weng17}. This
evolution sequence is in agreement with an accretion scenario where the
accretion column has a fan beam and a pencil beam pattern above and below the
critical luminosity \citep{basko76, becker12, mushtukov15, sartore15},
respectively. However, when taking a close look at the pulse profiles, we can
find another narrow peak that follows the second main peak at the late
stage of outburst, indicating the complicated geometry of NS magnetic field.

Investigating evolutions of isolated NSs spin period and its
derivative, we could obtain the key informations of NSs magnetic field and
characteristic age \citep[e.g. ][]{manchester77, lyne12, gao16, gao17}.
Alternatively, for BeXRBs, the observed NS spin frequency is modulated by the
orbital motion, which allows measurement of the orbital parameters of the
binary system \citep[e.g.,][]{li11, takagi16}. \cite{townsend17} tried
to measure the evolution of the pulsar period, taking into account both the
accretion induced spin-up effect \citep[described by the Ghosh \& Lamb
relation; ][]{ghosh79, wang81} and the modulation due to the binary orbital
motion. In this way, assuming that the spin-up rate is proportional to $L_{\rm
X}^{6/7}$, they tried to fit the orbital parameters; however, they could not
obtain an adequate fit (i.e. the fitting residuals globally increase since MJD
57675, Figure 6 in their paper). \cite{tsygankov17} took the variation of
bolometric correction into account, but the spin-up model still struggled to
the data with the sharp and complex features shown in the fitting residuals
(bottom panel of Figure 3 in their paper). Investigating the whole set of {\it
Swift} monitoring data, we found that the spin-up rate and the 0.6--10 keV flux
follows the power-law relation with an index of $0.84\pm0.02$, which is in
agreement with the predicted value of 6/7. But the relation deviates from the
power-law at the peak and the low luminosity (Figure 5 in Paper I).
\cite{townsend17} suggested that the variable accretion rate results in the
complex changes in spin-up rate, which cannot be well described by the
canonical spin-up model, and the higher order variations in the spin-up of SMC
X-3 are requested. We utilized seven frequency derivatives to model the spin
evolution and obtained the orbital parameters: the orbital period $P =
44.52\pm0.09$ days, the projected semi-major axis $asini = 194\pm1$ light
seconds, the longitude of periastron $\omega = 202\pm2 \degr$ and an
eccentricity of $e = 0.259\pm0.003$ (Weng et al. 2016, Paper I).

SMC X-3 was visited by {\it Nustar} at a high luminosity state during
the 2016-2017 giant outburst rise (on 2016 August 13), one {\it XMM-Newton}
(2016 October 14) and the second {\it Nustar} observations (2016 November 12)
were subsequently carried out at the intermediate luminosity level during the
decay of outburst (Table \ref{log}). The phase-averaged spectrum of {\it
XMM-Newton} has been reported in Paper I, and the {\it Nustar} observations
have been partially analyzed in \cite{tsygankov17}. Since both {\it XMM-Newton}
and {\it Nustar} have the large effective areas, high time resolution and
moderate energy resolution, we present a detailed pulse phase-resolved analysis
on these high quality data to explore the nature of accretion flows in this
extreme outburst. We describe the data reduction in the next section and
perform the timing and spectral analyses in Section 3. In order to
describe the evolution of the whole burst, we also consider the {\it Swift}
results presented in Paper I, in particular, at the low luminosity level.
Discussion and conclusion follow in Section 4.

\section{Data Reduction}

The European Photon Imaging Camera (EPIC) is the main science instrument of
{\it XMM-Newton}, and it is comprised of three X-ray CCD cameras, i.e. the pn
and two MOS cameras. The time resolution is as good as 0.03 ms for the EPIC-pn
timing mode data, and the energy resolution (FWHM) of EPIC is of $\sim
0.1-0.15$ keV at 6 keV \footnote{See {\it XMM-Newton} Users Handbook, \\
\protect\url{http://xmm-tools.cosmos.esa.int/external/xmm\_user\_support/documentation/uhb/}}.
Because the EPIC-MOS observation was performed in imaging mode and seriously
suffered from the pile-up effect, we only analyze the EPIC-pn data, which were
taken in timing mode. The observation in the first 4.5 ks is
contaminated by background flares, and therefore the data are excluded for the
following analysis. The rest data with a net exposure time of 28 ks
are reduced by using the Science Analysis System software (\textsc{sas})
version 16.0.0 with the standard filters: FLAG $=$ 0 and PATTERN $\leq$ 4. The
source region is centered in RAWX = 38 with a width of 18 pixels, while the
background region is centered in RAWX = 4 with a width of 2 pixels.


\begin{center}
\begin{table*}[t]
\caption{Log of observations}
\label{log}
\begin{tabular}{clccccc}\hline \hline

Obs Date & Observatory & ObsID &
Net Exposure (ksec) & Period (s) & $L_{\rm X}$\\ \hline

2016 Aug 13     & {\it Nustar}  & 90201035002 & 25 & 7.810645(1) & 9.05$\pm$0.03 \\
2016 Oct 14     & {\it XMM-Newton}  & 0793182901 & 28 & 7.772007(3) & 1.16$\pm$0.01 \\
2016 Nov 12     & {\it Nustar}  & 90201041002 & 42 & 7.771533(5) &
1.62$\pm$0.01\\ \hline
\tablecomments{$L_{\rm X}$: Phase-averaged unabsorbed X-ray luminosity in units
of 10$^{38}$ erg/s is calculated in 0.5--10 keV for {\it XMM-Newton} and 3--50
keV for {\it Nustar} data, assuming a distance to the source of 62.1 kpc
\citep{hilditch05, graczyk14, scowcroft16}. \label{log}}
\end{tabular}
\end{table*}
\end{center}

We extract the 0.3--12 keV source event and convert the observational time into
the solar system barycenter time system with the task
\texttt{barycen}. We subsequently adopt the epoch-folding method with
the ftool \texttt{efsearch} to determine the spin frequency and its
uncertainty by using a least squares fits of a Gaussian to the observed
$\chi^2$ value versus the period \citep{leahy87}. The best fitted period is
$7.772007\pm0.000003$ s, and the consistent value can be obtained from the
$Z_{1}^{2}$ test \citep{buccheri83} as well. The light curves are generated in
time bin size of 0.05 s, and are corrected for the telescope vignetting and
point-spread-function losses with the \textsc{sas} task \texttt{epiclccorr}.

We create the spectral response files using the \textsc{sas} task
\texttt{rmfgen} and \texttt{arfgen} for the subsequent phase-averaged and
phase-resolved spectral analysis. The spectra are rebinned with the task
\texttt{specgroup} to have at least 20 counts per bin and not to oversample the
energy resolution of EPIC-pn by more than a factor of 3. The spectra are fitted
in 0.5--10 keV with \textsc{xspec} 12.9.1 \citep{arnaud96}.

The {\it Nustar} observatory consists of two focusing instruments and two focal
plane modules (FPMA and FPMB), has a time resolution better than 1 ms and
an energy resolution of $\sim 0.3-0.4$ keV at 6 keV
\citep{harrison13}. The {\it Nustar} data are processed with the packages and
tools in \textsc{heasoft} version 6.21. The first {\it Nustar} observation was
carried out at the peak of outburst; therefore, the source photons are
extracted from a larger circular region with the radius of 120{\arcsec}. On the
other hand, a smaller region with an aperture radius of 60{\arcsec} is adopted
for the second {\it Nustar} observation due to relatively low count rate.
Meanwhile, the background photons are extracted from the source-free region.
The source events in 3--50 keV are extracted and applied for the barycenter
correction in order to calculate the spin period (Table \ref{log}). The spectra
and light curves are produced with the proper corrections using the
task \texttt{nuproducts}.


\section{Results}
\subsection{Timing analysis}

\begin{figure*}
\begin{center}
\includegraphics[scale=0.5]{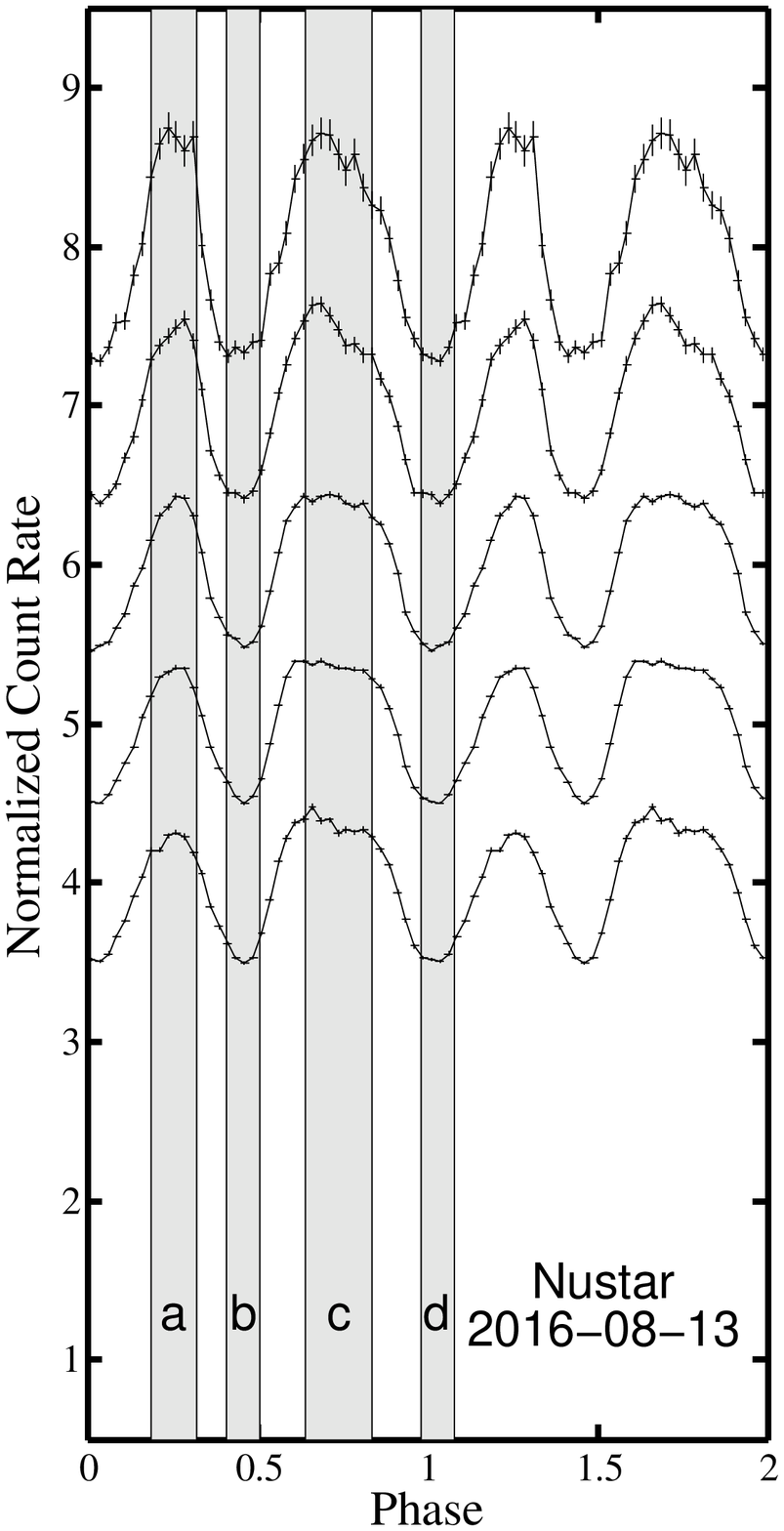}
\includegraphics[scale=0.5]{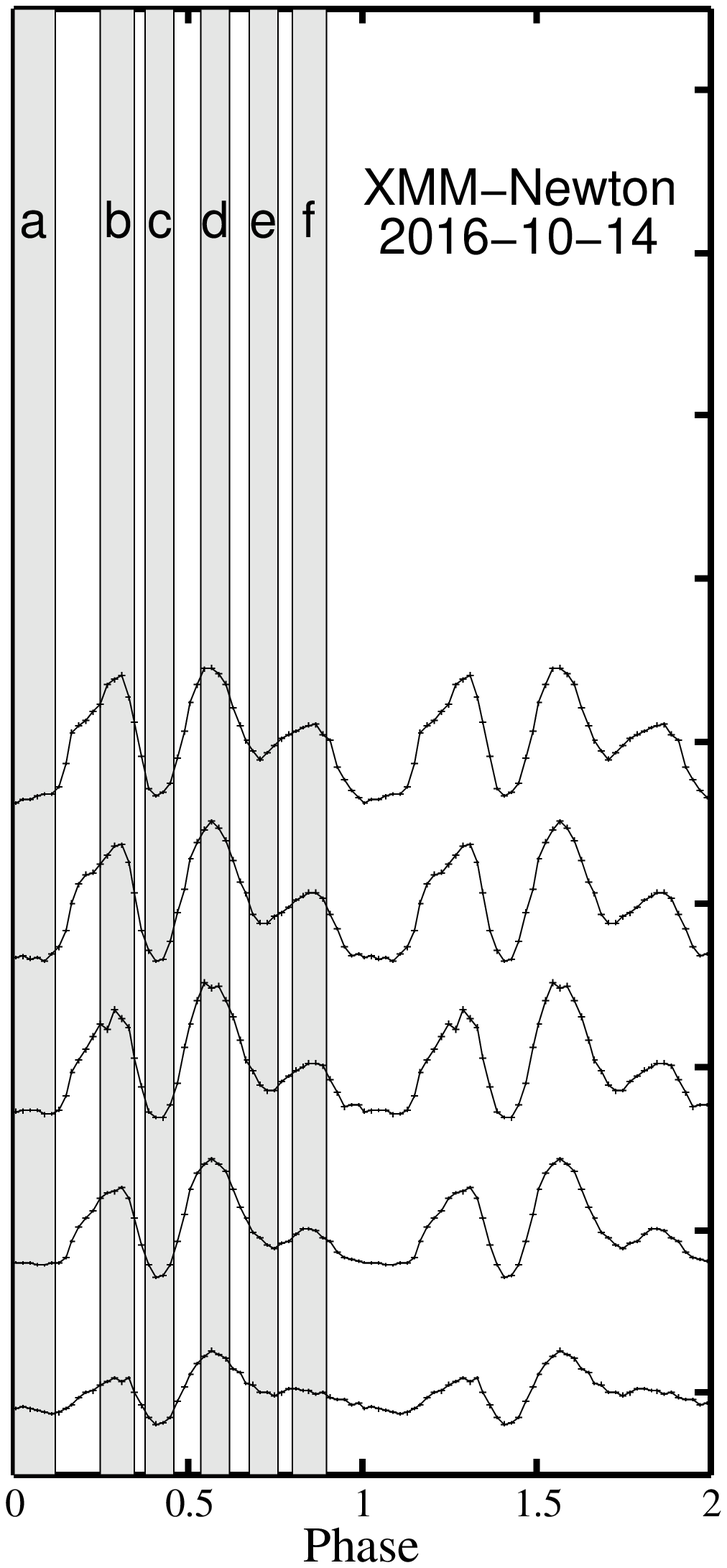}
\includegraphics[scale=0.5]{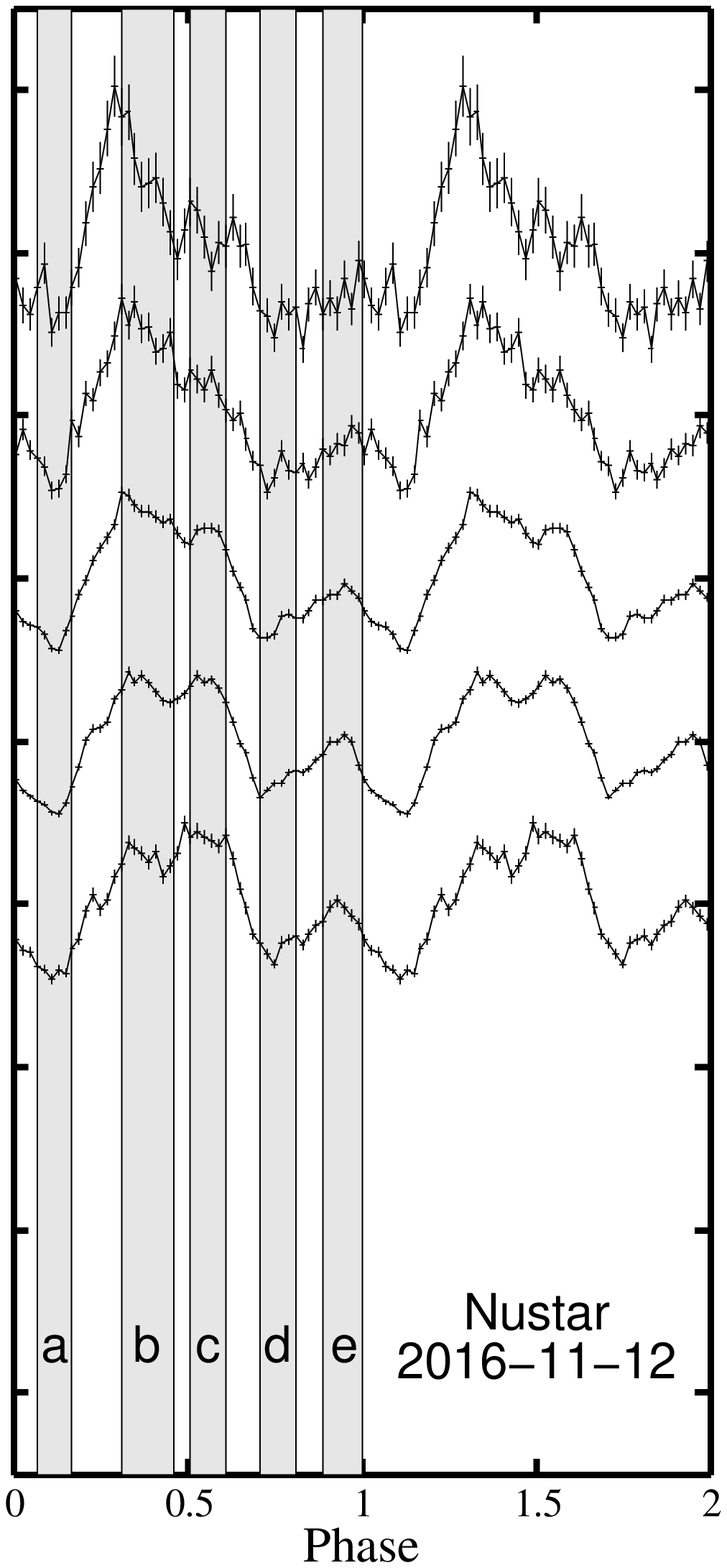}
\caption{Energy-dependent pulse profiles. For {\it Nustar} data (left and right
panels), from the bottom to the top, energy intervals correspond to 3--5 keV,
5--10 keV, 10--20 keV, 20--30 keV, and 30--50 keV, respectively. In the middle
panel, we show the {\it XMM-Newton} pulse profiles in 0.3--1 keV, 1--2 keV,
2--3 keV, 3--5 keV, and 5--10 keV from the bottom to the top, respectively. We
also arbitrarily shift up and plot two cycles of pulse profiles for clarity.
The grey boxes mark the pulse phase regions in which the phase-resolved spectra
are extracted and analyzed (Table \ref{spec}). \label{profile}}
\end{center}
\end{figure*}

In order to investigate the energy-dependent pulse profiles, we extract the
light curves in energy ranges of 0.3--1 keV, 1--2 keV, 2--3 keV, 3--5 keV, and
5--10 keV from {\it XMM-Newton} data, and in 3--5 keV, 5--10 keV, 10--20 keV,
20--30 keV, and 30--50 keV for {\it Nustar} data. Because the exposure time ($<
1$ day) is much shorter than the orbital period, the binary orbital modulation
in each observation is negligible, and the background-subtracted light curves
are folded over the observed spin period with a phase bin number of 50. The
evolution of pulse profiles is shown explicitly in Figure \ref{profile}. At the
high luminosity state, the typical fan beam pattern, i.e. the double-peak
profile, is exhibited in the first {\it Nustar} observation. The two peaks have
similar amplitudes and are separated by more than $\Delta\phi > 0.4$. At the
late stage, the two main peaks are converging ($\Delta\phi < 0.3$), and the
second peak disappears at the high energy ($> 20$ keV, right panel of Figure
\ref{profile}). In addition, another narrow peak (hereafter, we call it ``the
minor peak") emerges after the second peak.

\begin{figure}
\begin{center}
\includegraphics[scale=0.45]{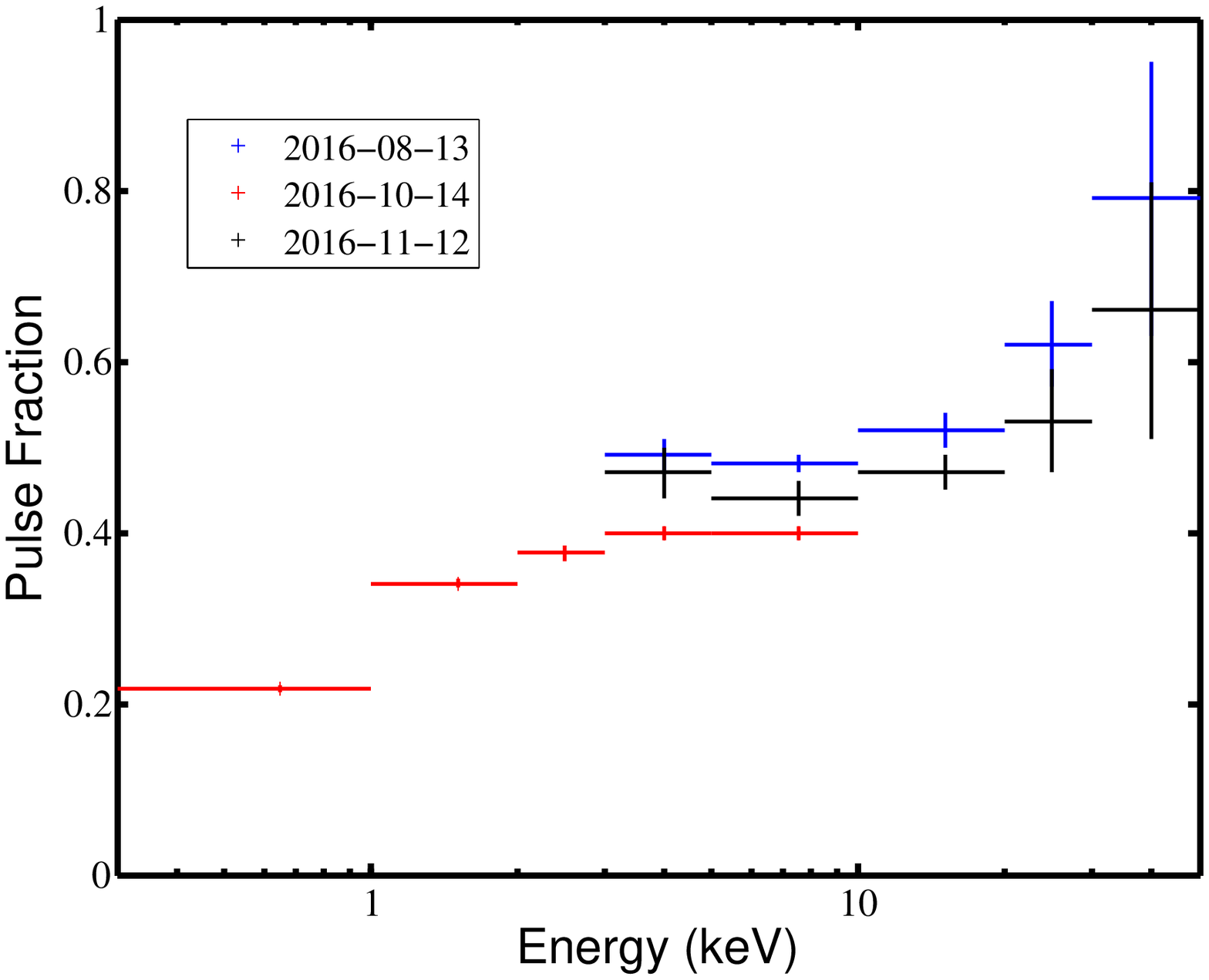}
\caption{Evolution of energy-dependent pulse fraction. \label{pf}}
\end{center}
\end{figure}

The pulse fraction is calculated as $PF=(M-N)/(M+N)$, where $M$ and $N$ are the
maximum and minimum count rates, respectively. The pulse fractions as a
function of photon energy are plotted in Figure \ref{pf}, in which the {\it
Nustar} data points are in agreement with those shown in Figure 5 of
\cite{tsygankov17}. The NS spin modulation amplitude increases with the photon
energy, while there is a hint of plateau in the range of 3-10 keV. The pulse
fraction below 1 keV ($PF = 0.218\pm0.004$) is significantly lower than those
detected beyond 1 keV.

\subsection{Spectroscopy}
\subsubsection{Model}

Following Paper I, we firstly employ the model consisting of a
black-body (BB) plus a power-law component, and a Gaussian line
(\texttt{tbabs*(bbodyrad+powerlaw+gauss)} in \textsc{xspec}) to fit the {\it
XMM-Newton} data (Table \ref{spec}). The phase-resolved spectral
analysis is carried out to explore the X-ray properties of two main peaks,
dips, and the minor peak if existed (the grey regions in Figure \ref{profile}).
There is no evidence for the change of absorption column density in different
pulse phases; thus, we fix the value to $0.138 \times 10^{22}$ cm$^{-2}$ in the
following modelling (Table \ref{spec}). The line width of iron emission
changes dramatically with the pulse phase, and becomes too narrow to be
resolved by {\it XMM-Newton} data in the phase range of $\sim 0.38-0.62$
(spectra c and d). In this case, we fix the line width to $\sigma =
0.05$ keV.

Meanwhile, an absorbed exponentially cutoff power-law model with an additional
fluorescent iron line is adopted for the {\it Nustar} spectral analysis. Due to
the lack of sensitivity below 3 keV, the {\it Nustar} data cannot provide a
substantial constraint on the neutral hydrogen column density. Thus, we fix the
hydrogen column ($N_{\rm H}$) at $0.138 \times 10^{22}$ cm$^{-2}$ derived from
the {\it XMM-Newton} spectral fitting. In \cite{tsygankov17}, an
additional BB component with a temperature $\sim 0.8$ keV was included in the
fitting to the first {\it Nustar} observation. However, this component was not
detected in the simultaneous {\it Swift}/XRT observations \citep[Paper I;
][]{tsygankov17, townsend17}, which are much more sensitive below 3 keV. We do
not know whether the thermal component detected by {\it Nustar} alone is real
or due to the calibration uncertainty. Nerveless, the BB component only
contributes less than 2\% of total flux in 3--50 keV, and it does not affect
the other parameters much. Therefore, we do not include this component in our
work.

For the second {\it Nustar} observation, the iron line is marginally detected
with a confidence level of $\sim 98\%$ according to $F$-test. In addition,
since the width of the fluorescent iron line becomes narrower than the energy
resolution of {\it Nustar} \citep{tsygankov17}, it was fixed at the value of
0.1 keV for the phase-averaged spectral analysis. But we do not involve the
Gaussian line in the phase-resolved spectral fitting because of low
significance level (Table \ref{spec}).

\subsubsection{Trends of Evolution}
The first {\it Nustar} observation was carried out at the outburst rise
with a high X-ray luminosity level. The significance of iron line in
phase-resolved spectra is greater than 99.99\% at least. The best-fit spectral
parameters at two main peaks are almost the same with $\Gamma \sim 1.0$ and
$E_{\rm cut} \sim 18$ keV. Alternatively, the radiation at two dips is harder,
that is, $\Gamma \sim 0.9$ and $E_{\rm cut} \sim 12.5$ keV.

A cool thermal component ($kT_{\rm BB} \sim 0.2$ keV) is evidently shown in the
{\it XMM-Newton} data obtained on 2016 October 14. The size of thermal
component ($R_{\rm BB} \sim 220$ km without the color correction) is
significantly larger than the radius of NS and varies with the pulse phases.
The emission at the second main peak ($\Gamma = 1.02$) is softer than the value
($\Gamma = 0.93$) at the first peak. The fitted photon indices for two dips are
$\Gamma = 1.17$ and 0.93, respectively. The source has the hardest spectrum
($\Gamma = 0.91$) at the minor peak. The significance of Gaussian line
in phase-resolved spectra is greater than 96\%.

Both the second {\it Nustar} and the {\it XMM-Newton} observations were
performed during the decay of outburst, and they share the similar spectral
evolution trend along the pulse phase. Compared with the {\it XMM-Newton} data,
the second {\it Nustar} data have a lower luminosity and harder spectral
profiles (Table \ref{spec}). There is no significant correlation found among
the best-fit parameters.


\begin{center}
\begin{table*}[t]
\tiny
\caption{BEST-FIT SPECTRAL PARAMETERS}
\label{spec}
\begin{tabular}{ccccccccccc}\hline \hline


Phase & $N_{\rm H}$ & $kT_{\rm BB}$  &$R_{\rm BB}$ &$\Gamma$ & $E_{\rm cut}$ &
$E_{\rm Gau}$ & $\sigma$ & $f_{\rm X}$ & $f_{\rm BB}/f_{\rm X}$ & $\chi^2/$dof \\
&$10^{22}$ cm$^{-2}$ &
 keV  & km & & keV & keV & keV &  & $\%$ &\\
\hline
&&&&2016 Aug 13 & {\it Nustar}&&&&\\
0--1                   & $0.138^{\dag}$ & \nodata  &  \nodata  &  $1.00_{-0.01}^{+0.01}$  &  $16.4_{-0.2}^{+0.2}$  &  $6.38_{-0.06}^{+0.06}$ & $0.34_{-0.07}^{+0.08}$  &  $1.95_{-0.01}^{+0.01}$  & \nodata &  1390.6/1291 \\
a: 0.19--0.32          & $0.138^{\dag}$ & \nodata  &  \nodata  &  $0.99_{-0.02}^{+0.02}$  &  $18.3_{-0.6}^{+0.6}$  &  $6.51_{-0.15}^{+0.16}$ & $0.30_{-0.17}^{+0.24}$  &  $2.71_{-0.02}^{+0.02}$  & \nodata &  1136.1/1161 \\
b: 0.41--0.51          & $0.138^{\dag}$ & \nodata  &  \nodata  &  $0.90_{-0.04}^{+0.04}$  &  $12.6_{-0.6}^{+0.6}$  &  $6.38_{-0.09}^{+0.09}$ & $0.19_{-0.10}^{+0.15}$  &  $0.95_{-0.01}^{+0.01}$  & \nodata &  741.7/749 \\
c: 0.64--0.84          & $0.138^{\dag}$ & \nodata  &  \nodata  &  $1.00_{-0.02}^{+0.02}$  &  $17.6_{-0.4}^{+0.5}$  &  $6.38_{-0.15}^{+0.17}$ & $0.28_{-0.21}^{+0.19}$  &  $2.77_{-0.02}^{+0.02}$  & \nodata &  1336.6/1284 \\
d: 0--0.08 \& 0.98--1  & $0.138^{\dag}$ & \nodata  &  \nodata  &  $0.91_{-0.04}^{+0.04}$  &  $12.4_{-0.6}^{+0.6}$  &  $6.38_{-0.12}^{+0.11}$ & $0.27_{-0.12}^{+0.17}$  &  $0.90_{-0.01}^{+0.01}$  & \nodata &  736.4/733 \\

\noalign{\smallskip} \hline \noalign{\smallskip}
&&&&2016 Oct 14  & {\it XMM-Newton}&&&&\\
0--1                   & $0.138_{-0.017}^{+0.018}$ & $0.19_{-0.01}^{+0.01}$  &  $220.6_{-42.4}^{+49.4}$  &  $0.99_{-0.01}^{+0.01}$  &  \nodata   &  $6.63_{-0.10}^{+0.13}$ & $0.34_{-0.10}^{+0.13}$  &  $0.25_{-0.01}^{+0.01}$ &  4.3 &  118.0/165 \\
a: 0--0.12             & $0.138^{\dag}$            & $0.18_{-0.01}^{+0.01}$  &  $234.3_{-18.3}^{+20.8}$  &  $1.17_{-0.03}^{+0.05}$  &  \nodata   &  $7.09_{-0.30}^{+0.46}$ & $0.70_{-0.24}^{+0.70}$  &  $0.17_{-0.01}^{+0.01}$ &  5.5 &  172.3/156 \\
b: 0.25--0.35          & $0.138^{\dag}$            & $0.21_{-0.01}^{+0.01}$  &  $171.8_{-15.1}^{+17.5}$  &  $0.93_{-0.02}^{+0.02}$  &  \nodata   &  $6.63_{-0.14}^{+0.15}$ & $0.26_{-0.17}^{+0.27}$  &  $0.32_{-0.01}^{+0.01}$ &  3.3 &  140.1/160 \\
c: 0.38--0.46          & $0.138^{\dag}$            & $0.17_{-0.01}^{+0.01}$  &  $252.9_{-21.6}^{+24.9}$  &  $1.03_{-0.02}^{+0.02}$  &  \nodata   &  $6.50_{-0.08}^{+0.09}$ & $0.05^{\dag}$           &  $0.18_{-0.01}^{+0.01}$ &  5.6 &  155.4/153 \\
d: 0.54--0.62          & $0.138^{\dag}$            & $0.18_{-0.01}^{+0.01}$  &  $211.1_{-27.0}^{+33.0}$  &  $1.02_{-0.02}^{+0.02}$  &  \nodata   &  $6.73_{-0.09}^{+0.09}$ & $0.05^{\dag}$           &  $0.36_{-0.01}^{+0.01}$ &  2.2 &  141.7/161 \\
e: 0.68--0.76          & $0.138^{\dag}$            & $0.19_{-0.01}^{+0.01}$  &  $247.1_{-18.1}^{+20.5}$  &  $0.97_{-0.02}^{+0.03}$  &  \nodata   &  $6.70_{-0.25}^{+0.30}$ & $0.44_{-0.26}^{+0.41}$  &  $0.23_{-0.01}^{+0.01}$ &  5.8 &  129.5/153 \\
f: 0.8--0.9            & $0.138^{\dag}$            & $0.19_{-0.01}^{+0.01}$  &  $227.1_{-15.7}^{+17.5}$  &  $0.91_{-0.02}^{+0.03}$  &  \nodata   &  $6.76_{-0.55}^{+0.55}$ & $0.78_{-0.52}^{+0.90}$  &  $0.27_{-0.01}^{+0.01}$ &  4.8 &  129.6/160 \\

\noalign{\smallskip} \hline \noalign{\smallskip}
&&&&2016 Nov 12 & {\it Nustar}&&&&\\
0--1                   & $0.138^{\dag}$ & \nodata  &  \nodata  &  $0.66_{-0.02}^{+0.02}$  &  $14.9_{-0.3}^{+0.3}$  &  $6.41_{-0.11}^{+0.11}$ & $0.1^{\dag}$  &  $0.35_{-0.01}^{+0.01}$ & \nodata  &  1152.8/1006 \\
a: 0.07--0.17          & $0.138^{\dag}$ & \nodata  &  \nodata  &  $0.69_{-0.07}^{+0.07}$  &  $16.3_{-1.5}^{+1.7}$  &  \nodata & \nodata  &  $0.21_{-0.01}^{+0.01}$ & \nodata &  502.4/532 \\
b: 0.31--0.46          & $0.138^{\dag}$ & \nodata  &  \nodata  &  $0.60_{-0.04}^{+0.04}$  &  $15.3_{-0.7}^{+0.7}$  &  \nodata & \nodata  &  $0.50_{-0.01}^{+0.01}$ & \nodata &  969.5/894 \\
c: 0.51--0.61          & $0.138^{\dag}$ & \nodata  &  \nodata  &  $0.72_{-0.05}^{+0.05}$  &  $13.7_{-0.8}^{+0.8}$  &  \nodata & \nodata  &  $0.43_{-0.01}^{+0.01}$ & \nodata &  798.1/729 \\
d: 0.71--0.81          & $0.138^{\dag}$ & \nodata  &  \nodata  &  $0.57_{-0.07}^{+0.07}$  &  $12.3_{-0.9}^{+1.0}$  &  \nodata & \nodata  &  $0.23_{-0.01}^{+0.01}$ & \nodata &  545.1/569 \\
e: 0.89--1             & $0.138^{\dag}$ & \nodata  &  \nodata  &
$0.57_{-0.05}^{+0.05}$ &  $11.8_{-0.7}^{+0.7}$  &  \nodata & \nodata  &
$0.30_{-0.01}^{+0.01}$ & \nodata & 699.8/670\\
 \hline
\tablecomments{$f_{\rm X}$: Unabsorbed X-ray flux in units of $10^{-9}$ erg
cm$^{-2}$ s$^{-1}$ is calculated in 0.5--10 keV for {\it XMM-Newton} and 3--50
keV for {\it Nustar} data. $f_{\rm BB}/f_{\rm X}$: The percentage of
total flux due to the thermal component. All errors are in the 90\% confidence
level. ${\dag}$: The value of parameter is fixed. }

\end{tabular}
\end{table*}
\end{center}


\section{Conclusions \& Discussions}

SMC X-3 experienced an extra long giant outburst, from 2016 August to 2017
March, with the peak X-ray luminosity of $\sim 10^{39}$ erg/s. A number of
observations were carried out to observe the outburst, including one {\it
XMM-Newton} and two {\it Nustar} observations \citep{weng16, weng17,
townsend17, tsygankov17}. Thanks to the large effective area, the good time and
energy resolution of {\it XMM-Newton} and {\it Nustar}, we perform a detailed
phase-resolved analysis on these data to acquire more information on the
extreme outburst of SMC X-3.

(1) At the early stage of outburst, SMC X-3 has a double-peak pulse profile in
the broadband (3--50 keV), and the spectra at two main peaks are very similar.
As the flux decays, the second main peak approaches the first one, and vanishes
above 20 keV (Figure \ref{profile}). That is, its spectrum becomes much softer
than the spectrum at the first peak (Table \ref{spec}). These results have been
partially reported in \cite{tsygankov17} and Paper I.

(2) After the middle of 2016 October, a minor peak emerging after the second
main peak was found in Paper I with the {\it Swift} monitoring data. Using the
high quality data from {\it XMM-Newton} and {\it Nustar}, we confirm this
feature, which however is not predicted in either the fan beam nor the pencil
beam pattern \citep[e.g., ][]{bildsten97}. It might be due to the non-dipole
component of NS magnetic field proposed by \cite{tsygankov17}. We also find
that the source has the hardest radiation at the minor peak.

(3) The first {\it Nustar} data taken at a high luminosity level
exhibit larger pulse fraction than the second observation taken at the decay of
outburst. For all three observations, we find that the pulse fraction
increases with the photon energy. The significantly low pulse fraction at 0.3-1
keV can be interpreted as the existence of an additional thermal component
located far away from the central NS, which was firstly detected in
Paper I. Investigating a sample of bright X-ray pulsars,
\cite{hickox04} suggested that the soft X-ray excess was a common feature in
these systems. Here, the spectral fitting to the {\it XMM-Newton} data yields a
ratio of the BB flux to the total flux $\sim 0.02-0.06$ (Table \ref{spec}) and
the size of BB component ($\sim 1000$ km with the color correction), which is a
little bit smaller than the corotation radius of SMC X-3 ($\sim 6000$ km, see
more details in Paper I). All these parameters are consistent with those of
bright pulsars discussed in \cite{hickox04}. For the pulsars with $L_{\rm X}
\geq 10^{38}$ erg/s, the reasonable explanation for the soft X-ray excess is
the reprocessing of hard X-rays by the inner region of truncated accretion disk
\citep{hickox04}.

In contrast, the pulse fraction in the range of 0.5--2 keV starts to exceed
that detected in 2-10 keV (Figure 1 in Paper I) when the pulse profile switches
from the double-peak to the single-peak at low luminosity (Figure 4 in Paper
I). That is, the pulse fraction has different correlations with the photon
energy beyond and below the critical luminosity.

(4) The relatively large line widths ($\sigma \sim 0.34$ keV) shown in the
first {\it Nustar} and {\it XMM-Newton} phase-averaged spectra could result
from the Keplerian rotation at a radius of $\sim 1000$ km, which agrees with
the size of cool thermal component (Paper I). Alternatively, the Gaussian line
becomes narrower than the resolution of {\it Nustar}, and it can be explained
in the way that the thermal component is pushed to a larger radius by the NS
magnetosphere at lower luminosity \citep{lamb73}. In the {\it
XMM-Newton} data fitting, the energy of iron line varies with the pulse-phase,
suggesting different ionization stages (i.e. the He-like and the H-like
ionization status) in different directions. Additionally, the small line widths
($< 0.15$ keV), derived from the {\it XMM-Newton} data fitting in the phase
range of $\sim 0.38-0.62$ (spectra c and d), indicate a complicated geometry of
accretion column.

\acknowledgments{} This research has made use of public data obtained from the
High Energy Astrophysics Science Archive Research Center, provided by NASA's
Goddard Space Flight Center. We thank the anonymous referee for the helpful
comments. We thank Jun-Xian Wang and Sergey Tsygankov for many valuable
discussions. This work is supported by the National Natural Science Foundation
of China under grants 11703014, 11673013, 11503027, 11373024, 11233003,
11433005, 11573023 and 11233001, National Program on Key Research and
Development Project (Grant No. 2016YFA0400803 and 2017YFA0402703). H.H.Z.
acknowledges support from the Natural Science Foundation from Jiangsu Province
of China (Grant No. BK20171028), and the University Science Research Project of
Jiangsu Province (17KJB160002). Q.R.Y. thanks support from the Special Research
Fund for the Doctoral Program of Higher Education (grant No. 20133207110006).


\begin{thebibliography}{}

\bibitem[Arnaud (1996)]{arnaud96} Arnaud, K. A. 1996, in ASP Conf. Ser. 101, Astronomical Data Analysis Software and Systems V, ed. G. H. Jacoby \&J. Barnes (San Francisco, CA: ASP), 17

\bibitem[Basko \& Sunyaev(1976)]{basko76} Basko, M. M., \& Sunyaev, R. A. 1976, MNRAS, 175, 395

\bibitem[Becker et al.(2012)]{becker12} Becker, P. A., Klochkov, D., Sch\"{o}nherr, G., et al. 2012, A\&A , 544, A123

\bibitem[Bildsten et al.(1997)]{bildsten97} Bildsten, L., Chakrabarty, D., Chiu, J., et al. 1997, ApJS, 113, 367


\bibitem[Buccheri et al.(1983)]{buccheri83} Buccheri, R., Bennett, K., Bignami, G. F., et al. 1983, A\&A, 128, 245

\bibitem[Casares et al.(2014)]{casares14} Casares, J., Negueruela, I. Ribo, M. et al. 2014, Nature 505, 378


\bibitem[Clark et al.(1978)]{clark78} Clark, G., Doxsey, R., Li F., Jernigan, J. G., \& van Paradijs, J. 1978, ApJ, 221, L37

%
%
%

\bibitem[Edge et al(2004)]{edge04} Edge, W. R. T., Coe, M. J., Corbet, R. H. D., Markwardt, C. B., \& Laycock, S. 2004, Astron. Telegram, 225, 1

%
\bibitem[Galache et al.(2008)]{galache08} Galache, J.L., Corbet, R.H.D., Coe, M.J., et al. 2008, ApJS, 177, 189

\bibitem[Gao et al.(2016)]{gao16} Gao, Z.-F., Li, X.-D., Wang, N., et al. 2016, MNRAS, 456, 55

\bibitem[Gao et al.(2017)]{gao17} Gao, Z.-F., Wang, N., Shan, H., et al. 2017, ApJ, 849, 19

\bibitem[Ghosh \& Lamb(1979)]{ghosh79} Ghosh, P., \& Lamb, F. K. 1979, ApJ, 234, 296

%
\bibitem[Graczyk et al.(2014)]{graczyk14} Graczyk, D., Pietrzy\'{n}ski, G., Thompson, I. B., et al. 2014, ApJ, 780, 59

\bibitem[Haberl et al.(1995)]{haberl95} Haberl F. 1995, A\&A, 296, 685

\bibitem[Haberl et al.(2008)]{haberl08} Haberl F., Eger P., Pietsch W., 2008, A\&A, 489, 327

\bibitem[Haberl \& Sturm(2016)]{haberl16} Haberl, F., \& Sturm, R. 2016, A\&A , 586, A81

\bibitem[Harrison et al.(2013)]{harrison13} Harrison, F. A., Craig, W. W., Christensen, F. E., et al. 2013, ApJ, 770, 103

\bibitem[Hickox et al.(2004)]{hickox04} Hickox, R. C., Narayan, R., \& Kallman, T. R. 2004, ApJ, 614, 881

\bibitem[Hilditch et al.(2005)]{hilditch05} Hilditch, R. W., Howarth, I. D., \& Harries, T. J. 2005, MNRAS, 357, 304

%
%
\bibitem[Lamb et al.(1973)]{lamb73} Lamb, F. K., Pethick, C. J., \& Pines, D. 1973, ApJ, 184, 271

\bibitem[Leahy(1987)]{leahy87} Leahy, D. A. 1987, A\&A, 180, 275

\bibitem[Li et al.(2011)]{li11} Li, J., Wang, W., \& Zhao, Y. H. 2011, MNRAS, 423, 2854

\bibitem[Lyne \& Graham-Smith(2012)]{lyne12} Lyne, A. G., \& Graham-Smith, F. 2012, Pulsar Astronomy (4th ed.; Cambridge: Cambridge Univ. Press)

\bibitem[Manchester \& Taylor(1977)]{manchester77} Manchester, R. N., \& Taylor, J. H. (ed.) 1977, Pulsars (San Francisco, CA: Freeman)

%
%
\bibitem[Mushtukov et al.(2015)]{mushtukov15} Mushtukov, A. A., Suleimanov, V. F., Tsygankov, S. S., \& Poutanen, J. 2015, MNRAS, 447, 1847

%

\bibitem[Postnov et al.(2017)]{postnov17} Postnov, K., Oskinova, L., \& Torrej\'{o}n, J. M. 2017, MNRAS, 465, L119

\bibitem[Reig(2011)]{reig11} Reig, P., 2011, Ap\&SS, 332, 1

%
\bibitem[Sartore et al.(2015)]{sartore15} Sartore, N., Jourdain, E., \& Roques, J.-P. 2015, ApJ, 806, 193

\bibitem[Scowcroft et al.(2016)]{scowcroft16} Scowcroft, V., Freedman, W. L., Madore, B. F., et al. 2016, ApJ, 816, 49


\bibitem[Takagi et al.(2016)]{takagi16} Takagi, T., Mihara, T., Sugizaki, M., Makishima, K., \& Morii M., 2016, PASJ, 68, S13

\bibitem[Townsend et al.(2017)]{townsend17} Townsend, L. J., Kennea, J. A., Coe, M. J., et al., 2017,
MNRAS, 471, 3878

\bibitem[Tsygankov et al.(2017)]{tsygankov17} Tsygankov, S. S., Doroshenko, V., Lutovinov, A. A., Mushtukov, A. A., \& Poutanen, J. 2017, A\&A, 605, A39
%

\bibitem[Wang(1981)]{wang81} Wang, Y.-M. 1981, A\&A, 102, 36

\bibitem[Weng et al.(2017)]{weng17} Weng, S.-S., Ge, M.-Y., Zhao, H.-H., Wang, W., Zhang, S.-N. 2017, Bian, W.-H., \& Yuan, Q.-R. 2017, ApJ, 843, 69 (Paper I)

\bibitem[Weng et al.(2016)]{weng16} Weng, S.-S., Ge, M.-Y., Zhao, H.-H., Wang, W., \& Zhang, S.-N. 2016, Astron. Telegram, 9731, 1

%
%

\end{thebibliography}
\end{document}